\documentclass{article}
\usepackage[latin9]{inputenc}
\usepackage{url}
\usepackage{amsmath}
\usepackage{amssymb}
\usepackage{graphicx}
\usepackage{esint}

\makeatletter

\providecommand{\tabularnewline}{\\}

\usepackage{ijcai15}

\usepackage{times}\usepackage{balance}

\newtheorem{definition}{Definition}
\newtheorem{theorem}{Theorem}

\title{Approximation of barter exchanges with cycle length constraints}
\author{Suiqian Luo \and Pingzhong Tang \and Chenggang Wu \and Jianyang Zeng\\
Institute for Interdisciplinary Information Sciences,\\
Tsinghua University\\
\url{luosq13@mails.tsinghua.edu.cn}, \url{kenshin@tsinghua.edu.cn}\\
\url{wuchenggang0316@gmail.com}, \url{zengjy321@tsinghua.edu.cn}}

\makeatother

\begin{document}
\maketitle 
\begin{abstract}
We explore the clearing problem in the barter exchange market. The
problem, described in the terminology of graph theory, is to find
a set of vertex-disjoint, length-restricted cycles that maximize the
total weight in a weighted digraph. The problem has previously been
shown to be \textsc{NP-hard}. We advance the understanding of this
problem by the following contributions.

We prove three constant inapproximability results for this problem.
For the weighted graphs, we prove that it is \textsc{NP-hard} to approximate
the clearing problem within a factor of $\tfrac{14}{13}$ under general
length constraints and within a factor of $\tfrac{434}{433}$ when
the cycle length is not longer than $3$. For the unweighted graphs,
we prove that this problem is \textsc{NP-hard} to approximate within
a factor of $\tfrac{698}{697}$.

For the unweighted graphs when the cycle length is not longer than
$3$, we design and implement two simple and practical algorithms.
Experiments on simulated data suggest that these algorithms yield
excellent performances.
\end{abstract}

\section{Introduction}

Over the past decade, kidney exchange and matching based market design
in general, have become one of the most appealing applications at
the interface of economics and computer science. In economics, designing
desirable matching mechanism has been a topic of intensive research,
ever since the seminal work on college admission and stable marriage
problem~\cite{gale1962,roth1992two}. In computer science and the
multiagent system community, designing and fielding efficient clearing
algorithms for such markets has been under close scrutiny lately~\cite{abraham07,Awasthi09,Ashlagi2010,Dickerson2012dynamic,Dickerson13,dickerson2014price,Li2014}.

In a typical kidney exchange system %
\footnote{See, for example, \url{http://www.unos.org/} %
}, a patient with renal disease teams up with a known but incompatible
donor. While the pair donate a kidney to help some other compatible
patient in the system, they obtain a compatible kidney in return.
Both patients receive a compatible kidney in the end, resulting in
social welfare improvement. Nowadays, kidney exchange serves as alternative
solution besides cadaver donations and has been fielded successfully
in a number of countries. For a comprehensive introduction of background,
refer to~\cite{roth2004kidney,roth2005pairwise,abraham07} and the
references therein.

In this paper, we explore the kidney exchange problem from a computational
perspective. The problem, described in the terminology of graph theory,
is to find a set of vertex-disjoint cycles that maximize the total
weight in a weighted digraph. Each vertex in the graph represents
a patient-donor pair and each arc represents compatibility between
the pairs, with the arc weight denoting the payoff by performing the
surgery. A cycle of length $L$ requires $2L$ people in simultaneous
surgeries. To lessen the logistical pressure imposed by simultaneous
surgeries, in practice, every cycle length is constrained to be less
than or equal to $3$~\cite{abraham07} %
\footnote{An alternative solution is to include ``chains'', which start with
a cadaver or altruistic donor that does not look for anything in return.
Chain plays an important role in current kidney exchange systems.
See, e.g.,~\cite{roth2004kidney,Dickerson2012chains,Ashlagi2012,dickerson2014price}.
We do not consider chains in the current paper. %
}. For $L\leq3$, Abraham et. al.~\cite{abraham07} show that the
kidney exchange problem is \textsc{NP-hard}. They demonstrate an effective
integer programming formulation with an advanced tree search algorithm
that can solve a graph with $10000$ nodes.

We further explore computational complexity of this problem, for both
$L\leq3$ and general $L$, in both weighted and unweighted graphs.
Our conclusion is that the problem, under various definitions, is
computationally hard to approximate. The seemingly-straightforward
algorithms we propose can give good solutions on simulated data. In
particular, we make the following contributions:

\subsection{Our contribution}
\begin{enumerate}
\item We prove that, for a weighted graph with general $L$, the kidney
exchange problem is \textsc{NP-hard} to approximate within a factor
of $\tfrac{14}{13}$. The inapproximability result is obtained via
a reduction from the inapproximability of \emph{maximum 3-variable
linear equations modular 2 satisfiability problem} (aka. \textsc{Max-3Lin-2})
by H{\aa}stad~\cite{Hastad01}.
\item We prove that, for a weighted graph with $L\leq3$, the kidney exchange
problem is \textsc{NP-hard} to approximate within a factor of $\tfrac{434}{433}$.
The proof is via a reduction from the \emph{bounded occurrence version
of the maximum 3-variable linear equations modular 2 satisfiability
problem} (aka. \textsc{Max-3Lin-2($3$)}).
\item We show that, for an unweighted graph with $L\leq3$, the kidney exchange
problem is \textsc{NP-hard} to approximate within a factor of $\frac{698}{697}$.
The proof is via a reduction of the \textsc{Maximum-3-Dimensional-Matching}
problem.
\item For the unweighted graph with $L=3$, we propose two algorithms which
are easy to analyse and implement. We implement the algorithms and
test them on simulated data. Both of them yield good experimental
performance on these data.
\end{enumerate}
For the three inapproximability results, these problems has previous
been known to be \textsc{APX-hard}~\cite{biro2006inapproximability},
i.e., there is no polynomial time approximation scheme for these problems
(i.e., hard to find a $1-\epsilon$ approximation ). We advance the
theoretical understandings on these problems by showing that it is
hard to find a constant approximation.

These results have important implications. First, our hardness results
complement the work by Abraham et. al.~\cite{abraham07} and can
serve as a justification of their choice of integer programming implementation
over approximation algorithm. Second, our proof techniques shed light
on the close relations between the kidney exchange problem and several
other landmark computational optimization problems, and thus can serve
as technical basis for proving similar results in the kidney exchange
literature. Third, the two simple and practical algorithms and experimental
results suggest that there exist fast, near-optimal algorithms in
practice.

\section{Preliminary}

\subsection{The kidney exchange problem}

We model kidney exchange as a directed weighted graph $\mathcal{G}=(V,E)(|V|=n,|E|=m)$,
where each vertex stands for a donor-patient pair and each edge represents
a possible one-way kidney exchange. A cycle in the graph serves as
a basic building block of the exchange outcome, where a patient receives
a kidney from the donor in the preceding vertex along the cycle. The
weight of each cycle is the sum of weights for each arc along the
cycle. Depending on the specific application scenarios, edges can
be both weighted~\cite{abraham07} or unweighted (with uniform weight
on each arc)~\cite{roth2005pairwise}. Our goal is then to find a
collection of vertex-disjoint cycles that maximize the total weight
in weighted graphs and the total size in unweighted graphs. In other
words, the goal is to find a clearing algorithm that maximizes social
welfare.

We denote the problem by \textsc{Max-Weight-$L$-Exchange} for weighted
graphs and\textsc{ Max-Size-$L$-Exchange} for unweighted graphs,
where $L$ is the restriction of the exchange cycle length.

\subsection{Gap problems and inapproximability}

We first recall basic definitions of gap version optimization problems.
For more information about the theory of gap problems and inapproximability
of optimization problems, see, e.g.,~\cite{vazirani2001}.

Let $\mathcal{A}$ be any maximization problem, then the gap version
of $\mathcal{A}$ with parameter $0<a<b\leq1$, denoted by \textsc{Gap}-$\mathcal{A}$-{[}a,b{]},
is the following decision problem: Given an instance $\mathcal{I}$
of $\mathcal{A}$, distinguish whether the optimal solution has fractional
size at least $b$ or less than $a$. When fractional size of the
optimal solution is between $a$ and $b$, any output suffices.

Clearly, if there exists a polynomial time $\frac{b}{a}$-approximation
algorithm of the original problem $\mathcal{A}$, we can distinguish
the two cases in polynomial time. Conversely, if the gap version problem
\textsc{Gap}-$\mathcal{A}$-{[}a,b{]} is \textsc{NP-hard}, then the
original problem $\mathcal{A}$ is \textsc{NP-hard} to approximate
within a factor of $\frac{b}{a}$.

Gap version of optimization problems plays a central role in proving
inapproximability results. It has been widely studied in the theory
community. Many landmark problems were shown to be \textsc{NP-hard}
to approximate.

\begin{definition} \textsc{Max-3Lin-$q$} is the following optimization
problem:

\textbf{Input:} $n$ variables $x_{1},x_{2},\ldots,x_{n}$ with range
$[q]=\{0,1,\ldots,q-1\}$, and a set of $m$ constraints $c_{1},c_{2},\ldots,c_{m}$,
where each $c_{i}$ is a linear equation modular $q$ with $3$ variables,
e.g., $x_{1}+x_{4}+x_{8}=1\mod q$\textbf{.}

\textbf{Output:} Find an assignment satisfying the maximum number
of constraints. \end{definition}

Following the monumental $\mathcal{PCP}$-theorem~\cite{arora1998,arora1998probabilistic},
H{\aa}stad has proved the following celebrated theorem~\cite{Hastad01},
which serves as the start point of many inapproximability results:

\begin{theorem}[\textbf{H{\aa}stad}] \label{thm:max3lin} \textsc{Gap-Max-3Lin-$q$}-$[\tfrac{1}{q}+\epsilon,1-\epsilon]$
is \textsc{NP-hard} for any small constant $\epsilon>0$. The theorem
also holds when $m=O(n)$ where the constant only depends on $\epsilon$.
\end{theorem}

One natural variation of the \textsc{Max-3Lin-2} problem is to restrict
the number of occurrence of each variable in all the equations, we
denote this problem by \textsc{Max-3Lin-2($t$)}. For $t=3$, the
problem has been shown to be \textsc{NP-hard} to approximate within
some constant factor~\cite{bermanysome}:

\begin{theorem} \label{thm:max3linbounded} The \textsc{Max-3Lin-2($3$)}
problem is \textsc{NP-hard} to approximate within a factor of $\tfrac{62}{61}$.
More precisely, the \textsc{Gap-Max-3Lin-2($3$)}-$[\tfrac{61}{62}+\epsilon,1-\epsilon]$
problem is \textsc{NP-hard} for any small constant $\epsilon>0$.
\end{theorem}

\textsc{3-Dimensional-Matching} is one of Karp's 21 \textsc{NP-complete}
problems~\cite{karp1972}. \textsc{Maximum-3-Dimensional-Matching}
is a generalization of maximum matching in bipartite graphs to maximum
hyperedge matching in tripartite graphs and also has been shown to
be \textsc{NP-hard} to approximate:

\begin{theorem}[\textbf{Berman and Karpinski 2003}] \label{thm:3dmatching}\textsc{
Gap-Maximum-3-Dimensional-Matching}-$[\tfrac{697}{700}+\epsilon,\tfrac{698}{700}-\epsilon]$
is \textsc{NP-hard} for any small constant $\epsilon>0$. \end{theorem}

\section{Max-Weight-L-Exchange}

In this section, we prove that the \textsc{Max-Weight-$L$-Exchange}
problem is \textsc{NP-hard} to approximate within factor $\tfrac{14}{13}$.
To our best knowledge, this problem is only known to be \textsc{APX-hard}.
Here, we give the first explicit constant factor inapproximability
result.

\begin{theorem} The \textsc{Max-Weight-$L$-Exchange} problem is
\textsc{NP-hard} to approximate within a factor $\tfrac{14}{13}$.
More precisely, the \textsc{Gap-Max-Weight-$L$-Exchange}-$[\tfrac{13}{14}+\epsilon,1-\epsilon]$
problem is \textsc{NP-hard}. \end{theorem}

\textbf{Proof}. We reduce from the \textsc{Gap-Max-3Lin-2}-$[\tfrac{1}{2}+\epsilon,1-\epsilon]$
problem.

\begin{figure}
\centering \includegraphics[width=0.9\linewidth]{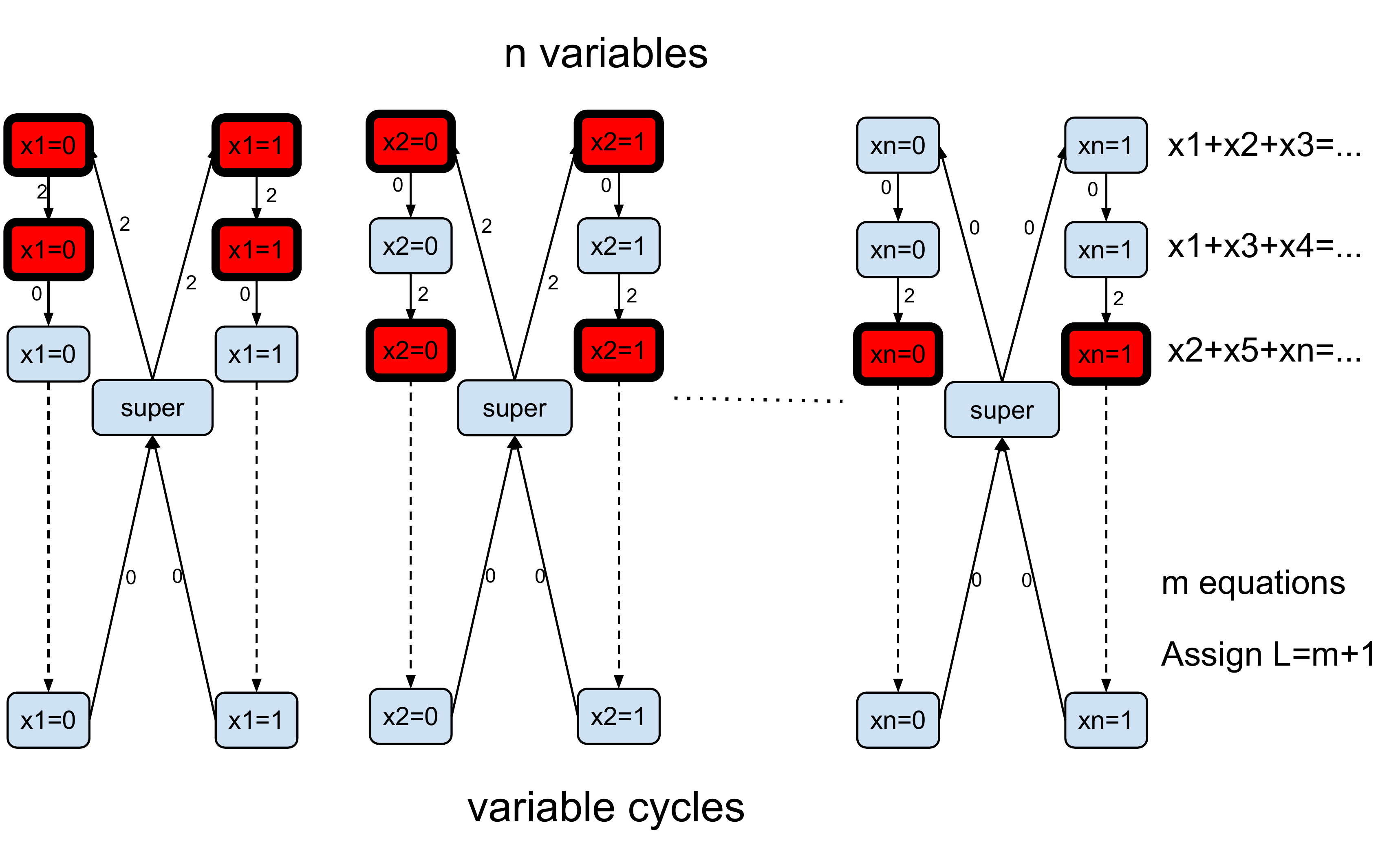}

\protect\caption{\label{fig:Construction for super node}The construction of \emph{variable
gadgets}}
\end{figure}

\textbf{Construction:} Given any \textsc{Max-3Lin-2} system with $n$
variables and $m$ equations (or constraints), we first construct
a \emph{variable gadget} for each variable $x_{i}$ with $2m+1$ nodes
as illustrated in Figure~\ref{fig:Construction for super node}.
The node in the center is named the \emph{super node} for $x_{i}$.
If $x_{i}$ appears in the $j$-th equation(as shown in boxes with
bold frames in Figure~\ref{fig:Construction for super node}), then
the arc entering the $j$-th row has weight $2$, otherwise the weight
will be $0$. We call the cycles passing through the super node \emph{variable
cycles}. The weight for any variable cycle associated with $x_{i}$
is $m_{i}$, where $m_{i}$ is the total number of equations that
variable $x_{i}$ appears.

\begin{figure}
\centering \includegraphics[width=0.7\linewidth]{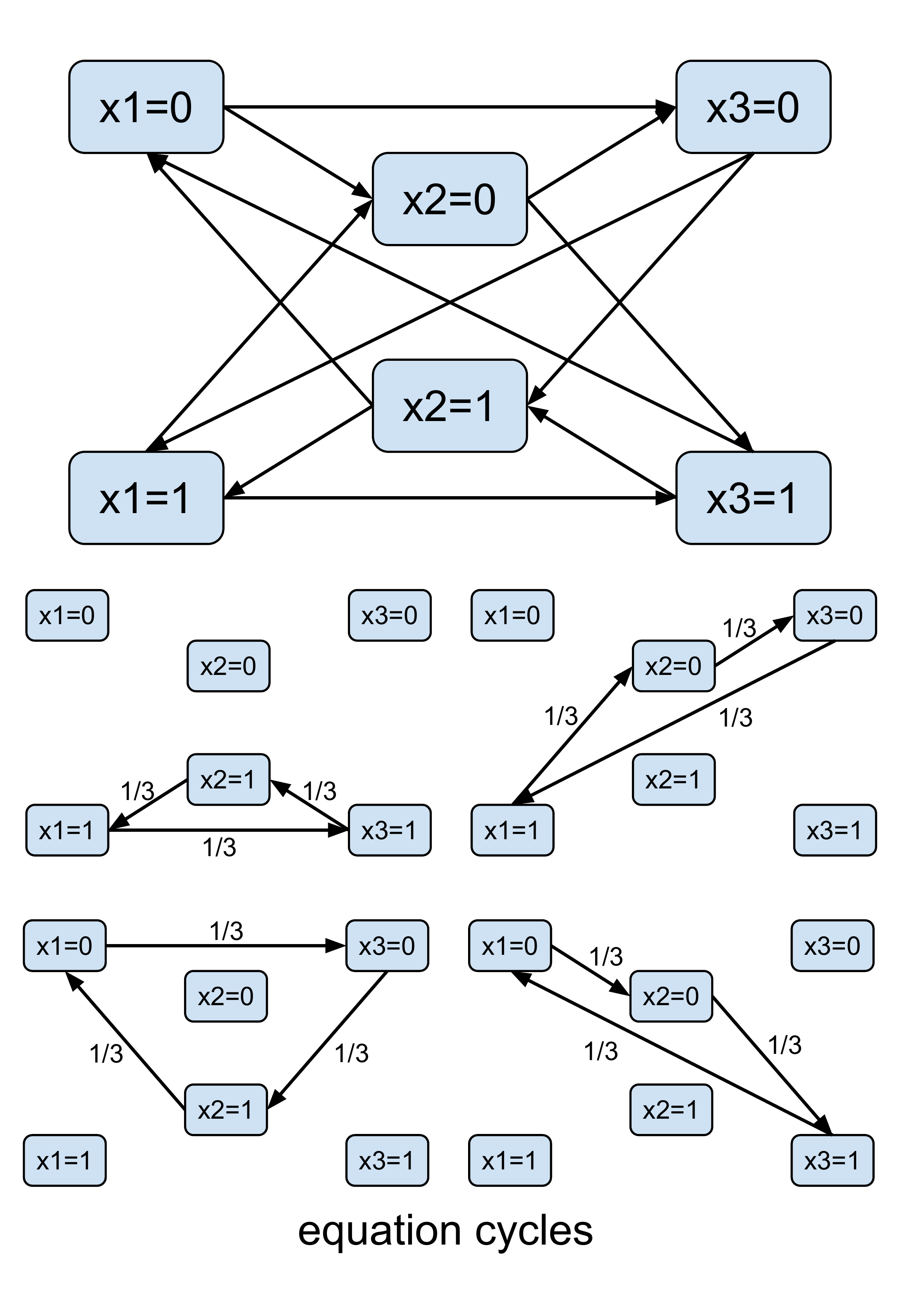}

\protect\caption{\label{fig:Construction for a constraint}The construction of \emph{equation
gadget} for $x_{1}+x_{2}+x_{3}\equiv1\mod2$}
\end{figure}

Second, for each equation, we use the six nodes in the corresponding
row to construct \emph{equation gadget}. Figure~\ref{fig:Construction for a constraint}
illustrates an example of constraint gadget construction for the equation
$x_{1}+x_{2}+x_{3}\equiv1\mod2$. If the equation is equal to $0$,
we can construct the graph symmetrically.

In each equation gadget, each arc is associated with weight $\frac{1}{3}$.
A property of the equation gadget is that there are exactly four cycles
with length $3$ in total for each equation and the weight of each
cycle is exactly $1$. There exists cycles which length is greater
than $3$ in this gadget, but the total weights is at most $2$. Each
cycle with length $3$ corresponds to one possible satisfiable assignment
of the equation. We call these cycles \emph{equation cycles}.

At last, we set $L=m+1$. The total number of vertices is $N=(2m+1)n=O(n^{2})$.
The reduction can be computed in polynomial time.

\textbf{Completeness:} We want to show that if there exists an assignment
$\sigma$ satisfying at least $(1-\epsilon)m$ constraints, we can
always construct a collection $C$ of cycles with a total weight at
least $6m+(1-\epsilon)m$. The proof is straightforward. For each
$x_{i}$, if $x_{i}$ is assigned with $1$, then we add the variable
cycle passing through nodes labeled with $x_{i}=0$; otherwise we
add the other variable cycle involving nodes labeled with $x_{i}=1$.
For each satisfiable constraint, we add the corresponding constraint
cycle in the corresponding constraint gadget. All the cycles in $C$
are vertex-disjoint. The sum of weight of variable cycles is $6m$,
and the sum of weight of equation cycles is at least $(1-\epsilon)m$.
Therefore the total weight is at least $6m+(1-\epsilon)m$.

\textbf{Soundness: } We need to show that if there is a collection
$C$ of cycles in the constructed graph with total weight $\geq6m+(\tfrac{1}{2}+\epsilon)m$,
we can resolve assignments to the variables such that at least $(\tfrac{1}{2}+\epsilon)m$
constraints are satisfied.

For each valid collection $C$ of cycles, if for each variable $x_{i}$,
there is exactly one variable cycle in $C$, then we call $C$ a \emph{good}
collection. Otherwise we call $C$ a \emph{bad} collection.

Our first claim is that for any bad collection $C$ of cycles, we
can always adjust $C$ to some good collection $C'$ without decreasing
the total weight. Suppose that neither of the variable cycles passing
through the super node for $x_{i}$ is included in $C$. Recall that
the number of appearance of $x_{i}$ is $m_{i}$. We know that there
will be at most $m_{i}$ cycles passing through the nodes associated
with the variable $x_{i}$. Adding the variable cycle will break up
at most $m_{i}$ equation cycles associated with $x_{i}$. We add
that variable cycle and delete all the broken equation cycles. As
the weight of the variable cycle is $m_{i}$, the total weight will
not decrease. By repeating this process, we will get a good collection.

Therefore we can assume that there is a good collection of cycles
with total weight $\geq6m+(\tfrac{1}{2}+\epsilon)m$. The contribution
of variable cycles is $3m$ exactly. Therefore there are at least
$(\tfrac{1}{2}+\epsilon)m$ equation cycles in $C$. We assign $x_{i}=0$
if the variable cycle passing through nodes on the right column is
included in $C$ and assign $x_{i}=1$ otherwise. As $C$ is a valid
collection of cycles, all the constraints with equation cycles in
$C$ are satisfied by the above assignments. Therefore at least $(\tfrac{1}{2}+\epsilon)m$
constraints are satisfied.

\textbf{Inapproximability ratio:} The total weight of cycles is at
most $7m$. Together with Theorem~\ref{thm:max3lin}, it is \textsc{NP-hard}
to distinguish whether the optimal solution is at least $\tfrac{6m+(1-\epsilon)m}{7m}=1-\epsilon'$
or at most $\frac{6m+(\frac{1}{2}+\epsilon)n}{7m}=\frac{13}{14}+\epsilon'$,
where $\epsilon'=\tfrac{1}{7}\epsilon$. Therefore the \textsc{Max-Weight-$L$-Exchange}
problem is \textsc{NP-hard} to approximate within $\tfrac{14}{13}$.

\section{Max-Weight-3-Exchange}

In this section, we prove that the \textsc{Max-Weight-3-Exchange}
problem is \textsc{NP-hard} to approximate within a constant factor:

\begin{theorem} The \textsc{Max-Weight-3-Exchange} problem is \textsc{NP-hard}
to approximate within $\tfrac{434}{433}$. More precisely, the \textsc{Gap-Max-Weight-3-Exchange}-$[\tfrac{433}{434}+\epsilon,1-\epsilon]$
is \textsc{NP-hard} for any small constant $\epsilon>0$. \end{theorem}

\textbf{Proof}. We reduce from a bounded occurrence version of the
\textsc{Max-3Lin-2($3$)} problem:

\textbf{Construction:} Given an instance of the \textsc{Max-3Lin-2($3$)}
with $n$ variables $x_{1},x_{2},\ldots,x_{n}$ and $m$ constraints
$c_{1},c_{2},\ldots,c_{m}$, where each constraint is a linear equation
modular 2 with 3 variables. Moreover, each variable $x_{i}$ appears
in at most 3 constraints. As shown in~\cite{bermanysome}, it is
\textsc{NP-hard} to distinguish whether the optimal solution satisfies
$\geq(1-\epsilon)m$ constraints or $\leq(\tfrac{61}{62}+\epsilon)m$
constraints for any constant $\epsilon>0$.

\begin{figure}
\centering \includegraphics[width=0.8\linewidth]{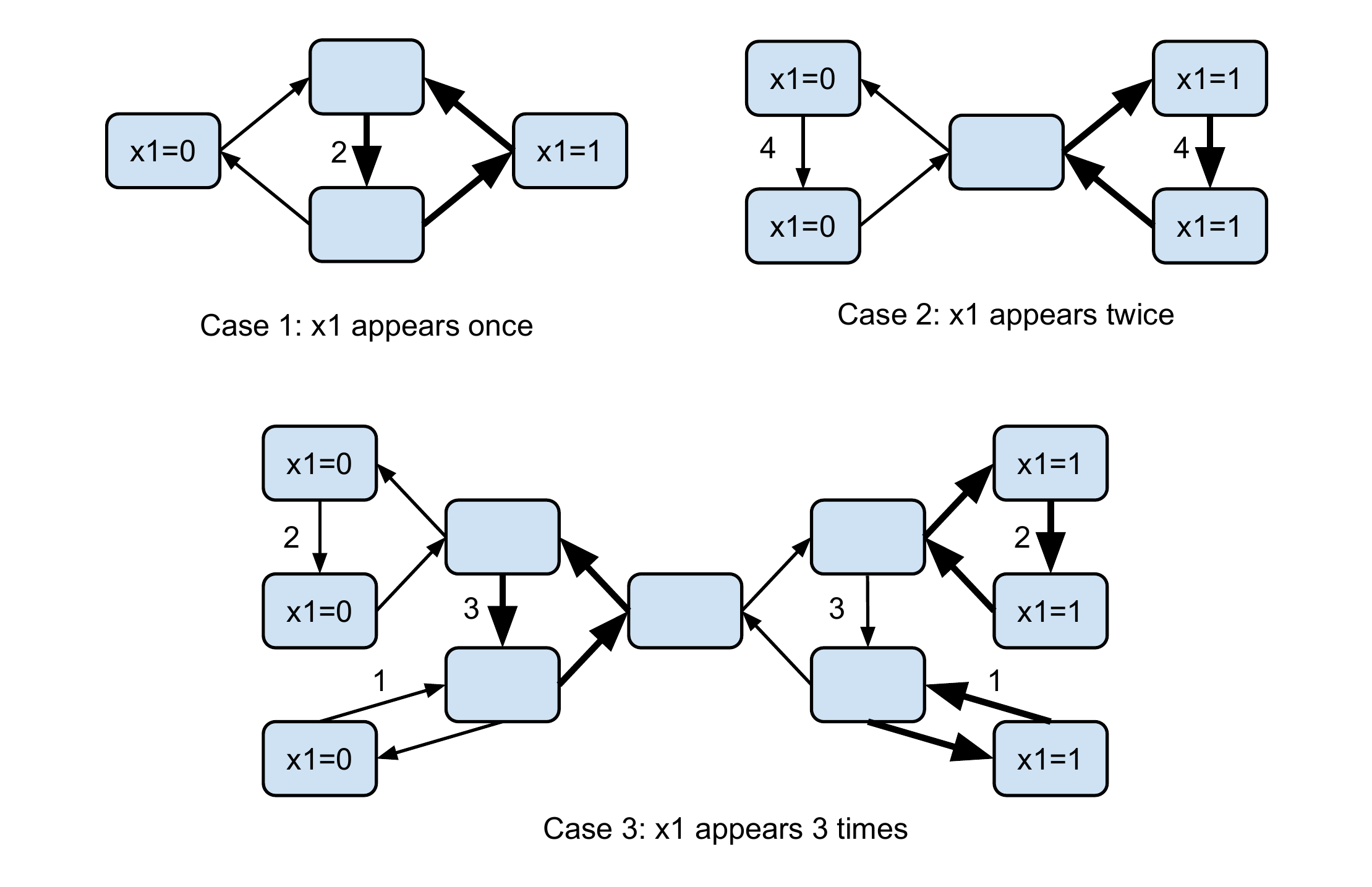}

\protect\caption{\label{fig:Construction for a variable}The construction of \emph{variable
gadget} for $x_{1}$}
\end{figure}

For each variable $x_{i}$, we construct a \emph{variable gadget}
as illustrated in Figure~\ref{fig:Construction for a variable} depending
on the number of appearances. Note that the total weight of cycles
with thin edges and the total weight of cycles with thick edges are
both exactly twice as the number of appearances of $x_{i}$.

For each equation constraint, we construct a \emph{constraint gadget}
involving the $6$ corresponding nodes. The constraint gadget is the
same as before, illustrated in Figure~\ref{fig:Construction for a constraint}.
Each of the four cycles with length $3$ has weight $1$ and the longer
cycles need not be considered since the constraint $L=3$. The reduction
can be computed in polynomial time.

\textbf{Completeness:} We show that if there is a good assignment
to the \textsc{Max-3Lin-2($3$)} instance with more than $(1-\epsilon)m$
satisfiable constraints, then we can construct a collection of cycles
$C$ as following: If $x_{i}=0$, then we add the cycles with thick
edges in the variable gadget, otherwise we add the cycles with thin
edges; for each constraint, if it is satisfied, then add the corresponding
cycle in the constraint gadget. As the sum of appearance of all variables
is $3m$, the weight contributed by the cycles from variable gadget
is $6m$. Moreover, there are at least $(1-\epsilon)m$ cycles from
the constraint gadgets. Therefore the total weight is at least $6m+(1-\epsilon)m$.

\textbf{Soundness:} We claim that if there is a collection $C$ of
cycles in the constructed graph with total weight $\geq6m+(\tfrac{61}{62}+\epsilon)m$,
we can find assignments to the variables such that $(\tfrac{61}{62}+\epsilon)m$
constraints are satisfied.

For any collection $C$ of cycles, if for all variable gadgets, either
all the cycles with thin edges are in $C$ or all the cycles with
thick edges are in $C$, then we call $C$ a \emph{good} collection.
Otherwise we say $C$ is \emph{bad}.

We claim that for any bad collection $C$, we can adjust $C$ to a
good collection $C'$, while the total weight will not decrease. We
assume that there is a good collection of cycles with weight $\geq6m+(\tfrac{61}{62}+\epsilon)m$.
As the contribution by cycles from variable gadgets is exactly $6m$,
there are at least $(\tfrac{61}{62}+\epsilon)m$ cycles from constraint
gadgets in $C$. We construct an assignment as following: for each
$x_{i}$, if all the thin edges are in $C$, then we assign $x_{i}=1$;
otherwise we assign $x_{i}=0$. For each cycle from constraint gadgets
in $C$, the corresponding constraint is satisfiable under the above
assignment. So at least $(\tfrac{61}{62}+\epsilon)m$ constraints
are satisfied.

\textbf{Inapproximability ratio:} The total weight is at most $7m$.
Together with Theorem~\ref{thm:max3linbounded}, it is \textsc{NP-hard}
to distinguish whether the optimal solution is at least $\tfrac{6m+(1-\epsilon)m}{7m}=1-\epsilon'$
or at most $\tfrac{6m+(\frac{61}{62}+\epsilon)m}{7m}=\tfrac{433}{434}+\epsilon'$,
where $\epsilon'=\tfrac{\epsilon}{7}$. Therefore the \textsc{Max-Weight-3-Exchange}
problem is \textsc{NP-hard} to approximate within $\tfrac{434}{433}$.

\section{Max-Size-3-Exchange}

In this section we prove an inapproximability result for the \textsc{Max-Size-3-Exchange}
problem. The unweighted exchange problem is a special case of the
weighted exchange problem with equal weights on each edge, therefore
the weighted exchange problem is even harder. Thus this inapproximability
result also holds for \textsc{Max-Weight-3-Exchange}.

\begin{theorem} The \textsc{Max-Size-3-Exchange} problem is \textsc{NP-hard}
to approximate within $\tfrac{698}{697}$. More precisely, the \textsc{Gap-Max-Size-3-Exchange}-$[\tfrac{697}{700}+\epsilon,\tfrac{698}{700}-\epsilon]$
is \textsc{NP-hard} for any small constant $\epsilon>0$. \end{theorem}

\textbf{Proof}. We reduce from the \textsc{Gap-Maximum-3-Dimensional-Matching}
problem.

\textbf{Construction:} In~\cite{berman2003improved}, a family of
\textsc{3-Dimensional-Matching} instances have been constructed, where
$|X|=|Y|=|Z|=100k$ for some integer $k$, with $m=200k$ triples,
and each element in $X\cup Y\cup Z$ appears in exactly 2 triples.
It is \textsc{NP-hard} to distinguish whether the size of the maximum
matching is at least $(98-\epsilon)k$ or at most $(97+\epsilon)k$.

\begin{figure}
\centering \includegraphics[width=0.8\linewidth]{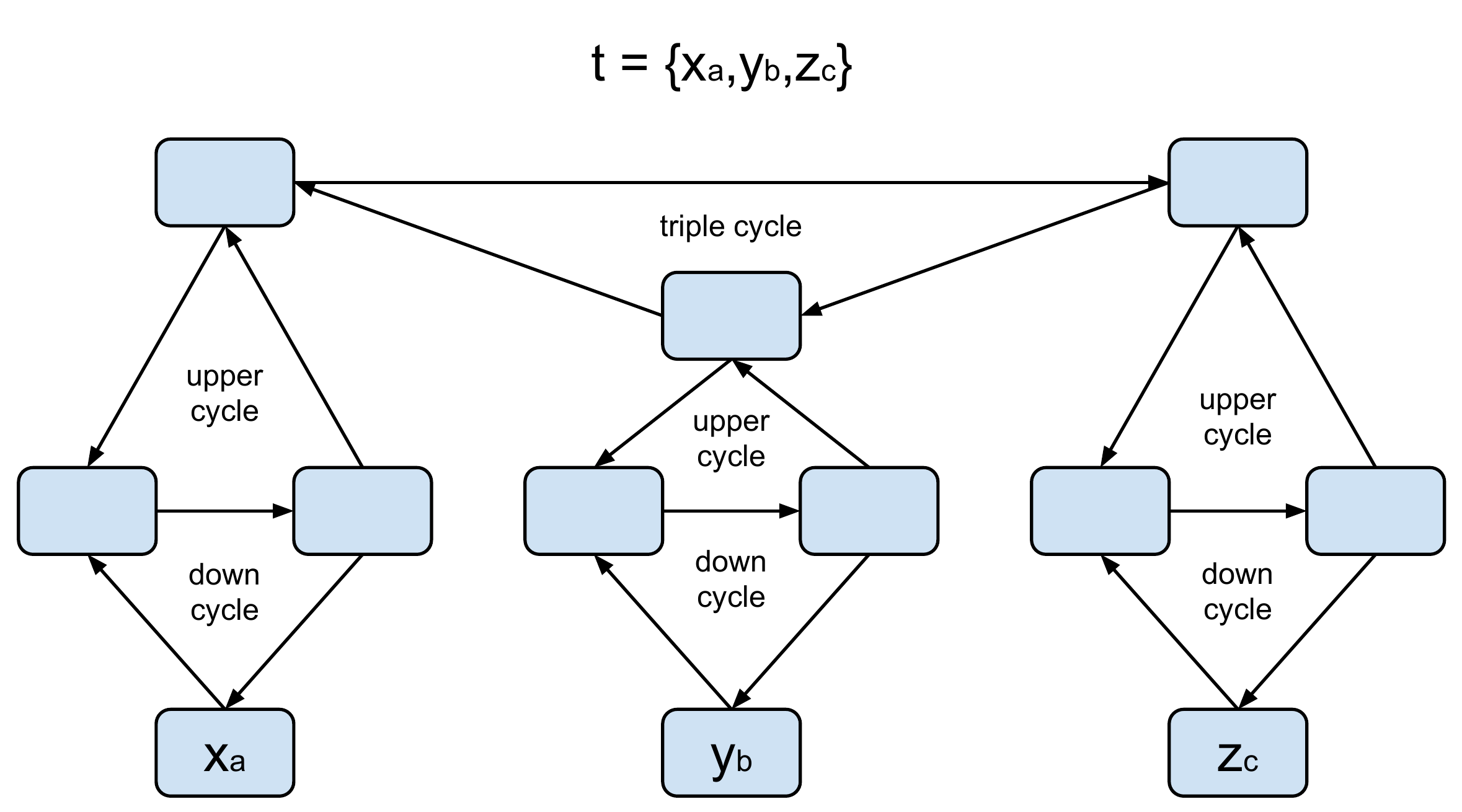}

\protect\caption{\label{fig:Construction for a triple}The construction for a triple
$t=\{x_{a},y_{b},z_{c}\}$}
\end{figure}

Given such an instance, we construct a graph $G$ as followings: for
each element in $X\cup Y\cup Z$, we have a node for the element with
the same label; for each triple $t=\{x_{a},y_{b},z_{c}\}\in T$, we
add a gadget as shown in Figure~\ref{fig:Construction for a triple}.
There are 7 cycles in the gadget, we call the three cycles at the
bottom \emph{down cycles}, the three cycles in the middle \emph{upper
cycles}, and the cycle on the top \emph{triple cycle}. All cycles
has uniform weight. The reduction can be computed in polynomial time.

\textbf{Completeness:} If there is a matching $T'\subseteq T$ with
size $\geq(98-\epsilon)k$, we construct a collection $C$ of cycles
as followings: for each triple $t=\{x_{a},y_{b},z_{c}\}$, if $t\in T'$,
then we add the \emph{triple cycle} and the three \emph{down cycles}
to $C$. Otherwise we add the three \emph{upper cycles} into $C$.
As $T'$ is a matching, all the cycles in $C$ are vertex disjoint.
The total number of cycles in $C$ is at least $200k*3+(98-\epsilon)k=(698-\epsilon)k$.

\textbf{Soundness:} If there is a collection $C$ of disjoint cycles
of size at least $(697+\epsilon)k$, we construct a matching $T'\subseteq T$
with size at least $(97+\epsilon)k$.

For the gadget for any triple, if the corresponding triple cycle is
in $C$, then the three upper cycles are not in $C$; if any of the
down cycles is not in $C$, then there are at most $3$ cycles in
$C$ within this gadget, therefore we can replace these cycles with
the three upper cycles, and the number of cycles will not decrease.
If the triple cycle is not in $C$, then we can just choose the 3
upper cycles, making no effects on other gadgets, and the number of
cycles will not decrease. After the adjusting processes, within each
gadget, either the three upper cycles are chosen, or the triple cycle
and the three down cycles are chosen. Therefore there are at least
$(97+\epsilon)k$ triple cycles in $C$. All the corresponding triples
are pairwise disjoint, otherwise the down cycles will not be disjoint.
So there is a matching of size at least $(97+\epsilon)k$.

\textbf{Inapproximability ratio:} The total number of cycles is at
most $700k$. Together with Theorem~\ref{thm:3dmatching}, it is
\textsc{NP-hard} to distinguish whether the optimal solution is at
least $\tfrac{(698-\epsilon)k}{700k}=\tfrac{698}{700}-\epsilon'$
or at most $\tfrac{(697+\epsilon)k}{700k}=\tfrac{697}{700}+\epsilon'$,
where $\epsilon'=\tfrac{\epsilon}{700}$. Therefore it is \textsc{NP-hard}
to approximate the \textsc{Max-Size-3-Exchange} problem within $\tfrac{698}{697}$.

\section{Algorithms for Max-Size-3-Exchange}

In this section, we mainly focus on the \textsc{Max-Size-3-Exchange}
problem. Previously, we have shown that it is \textsc{NP-hard} to
approximate within a factor of $\frac{698}{697}$. Now we propose
two simple and practical algorithms.

\subsection{Algorithm 1 via greedy search}

We present the basic greedy algorithm first. For the \textsc{Max-Size-3-Exchange}
problem, we figure out all the cycles with length $2$ or $3$ and
initialize our solution to be an empty set. We then pick up the cycles
with length $2$ or $3$ one by one and check whether it is available
to be appended in the solution. After this greedy process, the solution
is the final result of the algorithm.

\begin{theorem} \label{thm:approximation ratio} The approximation
ratio of the basic greedy algorithm is $3$. \end{theorem}

\begin{figure}
\centering \includegraphics[width=0.8\linewidth]{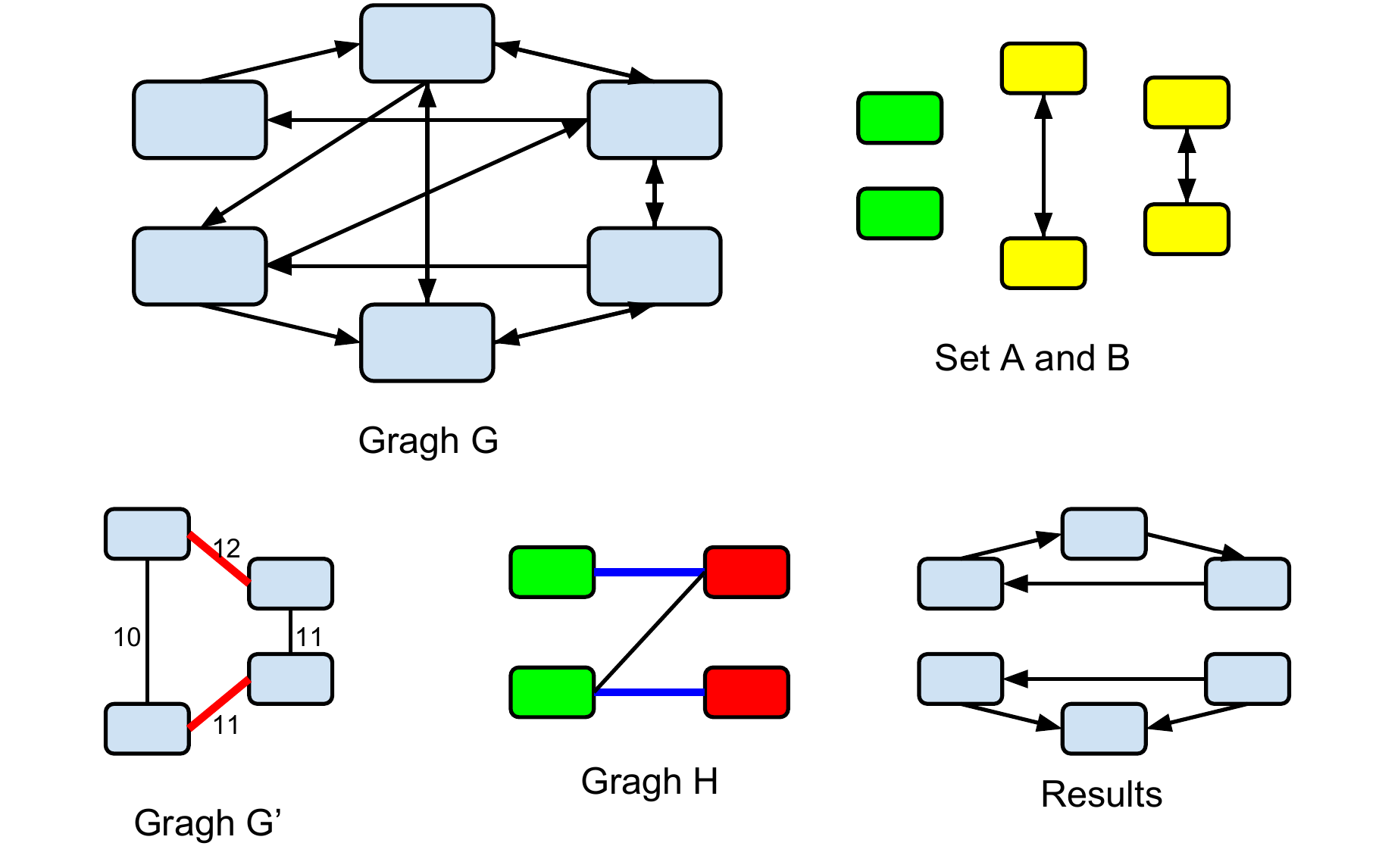}

\protect\caption{\label{fig:An example of approximation algorithm 2}An example of
algorithm 2}
\end{figure}

\textbf{Proof}. For any given graph $G$ in the \textsc{Max-Size-3-Exchange}
problem, suppose the optimal solution is $T$. Assume that there are
$d$ and $t$ cycles with length $2$ and $3$ respectively in the
optimal solution so that the size of solution is $|T|=2\times d+3\times t$.
We denote the approximation result by $T'$. It is clear that for
any cycles with length $2$ or $3$, there is at least one node which
is in $T'$. As a result, we have $|T'|\geqslant d+t$ and 
\[
\frac{|T|}{|T'|}\leq\frac{2\times d+3\times t}{d+t}\leq3.
\]

So the approximation ratio is $3$.

We now add an important heuristics that improves the performance of
the basic greedy algorithm. We consider the degree of each node in
the graph. Intuitively, the smaller the degree of node is, the more
difficult it is to be chosen in a cycle. The strategy is that we pick
the cycles with length $3$ first and pick up the cycles with smaller
degrees in priority by ordering the nodes according to the number
of degrees.

Given a graph $G$, calculate the number of in-degree and out-degree
for each node. Sort the order of nodes increasingly by $\#(indegree)\times\#(outdegree)$.
Find out all the cycles with length $3$ in $G$ and sort them with
lexicographical order. From the smallest number to the biggest number,
pick up the cycle if it is available, and delete the nodes in $G$.
After that, using the same method, find out all the cycles with length
$2$ in the remaining graph $G$, then sort them and pick up one by
one if possible. At last, output all the cycles which we have picked
up as final results. In the experiments section, we will implement
the algorithm above and test its performance using simulated data.

\subsection{Algorithm 2 via maximum matching}

Based on the fact that given a graph, the optimal solution of the
\textsc{Max-Size-2-Exchange} is close to the solution of \textsc{Max-Size-Exchange}
problem~\cite{abraham07}, we propose a straightforward method to
improve the results of \textsc{Max-Size-2-Exchange}. The idea is that,
after we figure out the maximum matching, we use the remaining nodes
and the matching edges to constitute cycles with length $3$ as far
as possible. For this purpose, we create a weighted graph and compute
the maximum matching for the second time in order to improve the opportunity
of using the remaining nodes.

\begin{figure}
\centering \includegraphics[width=0.8\linewidth]{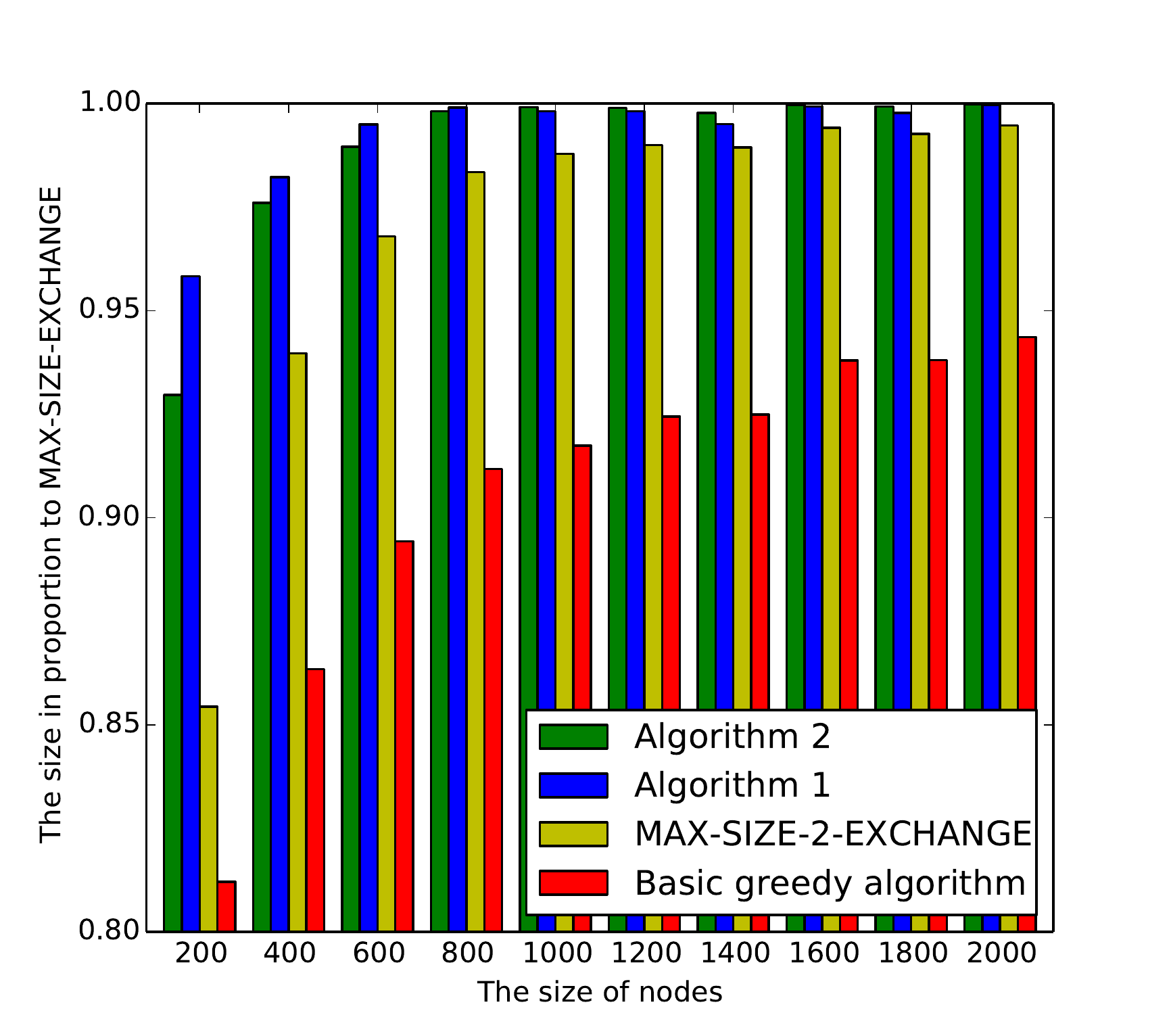}

\protect\caption{\label{fig:The experiment results of U.S. data}The experiment results
of the US data}
\end{figure}

\begin{figure}
\centering \includegraphics[width=0.8\linewidth]{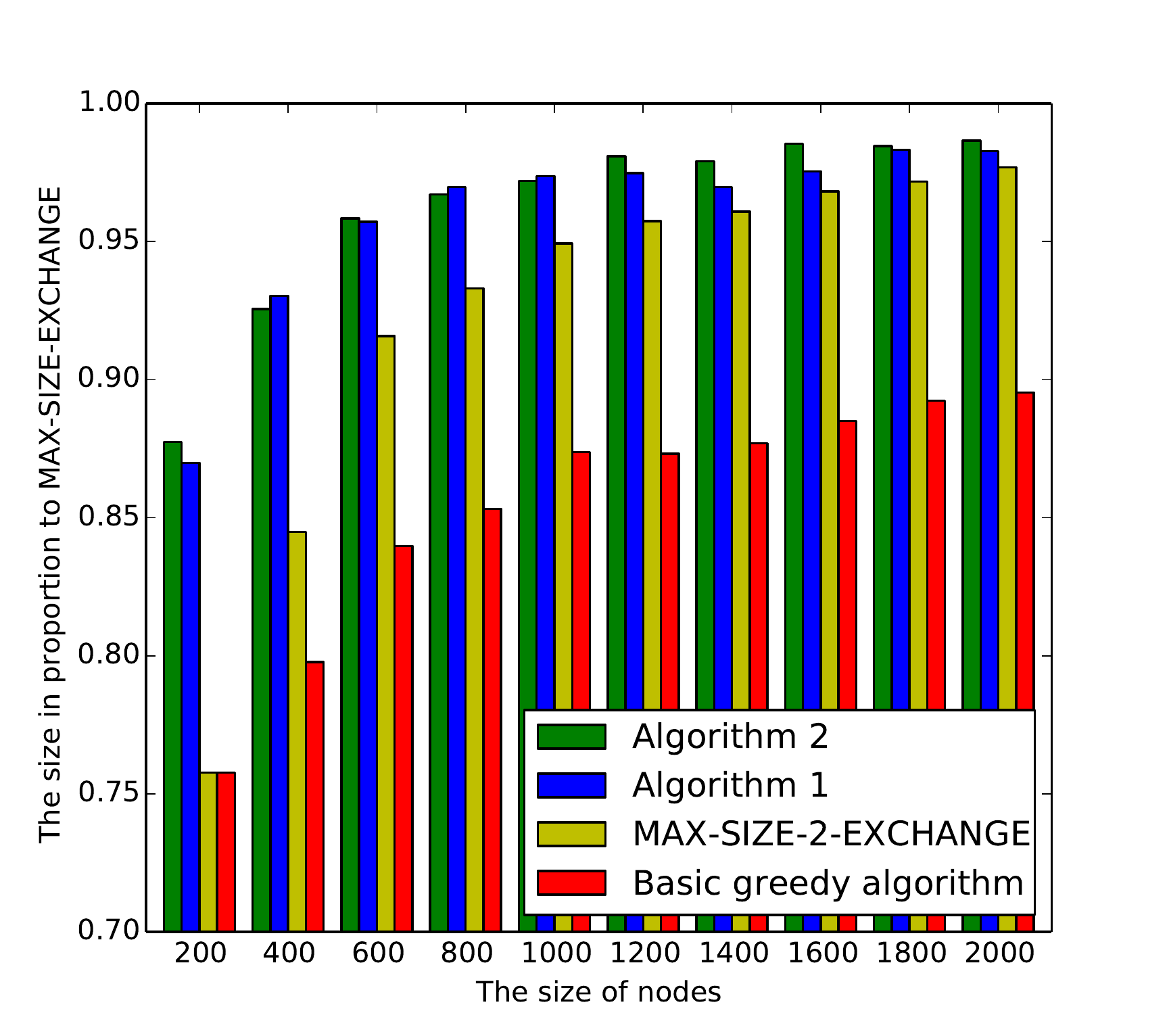}

\protect\caption{\label{fig:The experiment results of Chinese data}The experiment
results of the China data}
\end{figure}

Given a graph $G$, find the maximum matching in the bi-directional
edge of $G$ first. Suppose $A$ is the set of nodes which are not
in the maximum matching and $B$ is the nodes which are in the maximum
matching. As shown in Figure~\ref{fig:An example of approximation algorithm 2},
the colours of set $A$ and $B$ are green and yellow respectively.
For each node $u$ in $B$, we construct a new node $u'$ in $G'$.
For any cycle $(u,v)$ in $B$, we add a edge $(u',v')$ into $G'$
with a constant weight $w$ ($w=10$ in Figure~\ref{fig:An example of approximation algorithm 2}).
For any node $a$ in $A$, if node $a,u,v$ can constitute a cycle,
then add a edge $(u',v')$ with weight $1$. The weights of the same
edge will be added up. Then find the maximum matching of $G'$ as
illustrated by the red edges. Assume the matching is $M$. Construct
a bipartite graph $H$. The nodes in the left side stand for the nodes
of $A$ while the right side represent the edges in $M$. If the node
and the edge can constitute a cycle in the original graph $G$, then
add the edge in $H$. At last, find the maximum matching of $H$ as
shown by the blue edges. For any matching edge in $H$, output the
corresponding cycle with length $3$ in $G$. For any node in the
right side of $H$ but not in the matching edge, output the corresponding
cycle with length $2$ if exists.

Both algorithms 1 and 2 are easy to implement and run in polynomial
time. We will show in the next section that the two simple algorithms
yield good experimental performance.

\section{Experimental results}

In this section, we implement the algorithms proposed in the previous
section and compare their performance both in terms of running time
and solution quality.

\subsection{Experiments setup}

All our experiments are performed in Linux (openSUSE 13.1), using
a PC with four 3.2GHz Intel i5-3470 processors, and 4GB of RAM. Our
experimental data is carefully simulated based on the statistics of
US and China populations. We simulate the US data according to UNOS
waiting list and living donors, and simulate China data based on the
transplant researches~\cite{Tan2006Tissue,Tu2005The}.

Since it is \textsc{NP-hard} to figure out the optimal solution in
polynomial time, we use the following method to analyse the experimental
performance. Both \textsc{Max-Size-2-Exchange} and \textsc{Max-Size-Exchange}
can be solved in polynomial time using the maximum matching technique~\cite{abraham07}.
For any given graph, since the size of \textsc{Max-Size-3-Exchange}
will not be smaller than the \textsc{Max-Size-2-Exchange} and the
size of \textsc{Max-Size-Exchange} will be the largest among them,
the results of \textsc{Max-Size-2-Exchange} and \textsc{Max-Size-Exchange}
can be considered as the lower bound and the upper bound of \textsc{Max-Size-3-Exchange}
problem respectively. We compare the solution of our algorithms with
these bounds in order to analyse the performance.

For a particular size of nodes, we randomly generate $10$ copies
of the simulated graph and calculate the average of solutions of each
algorithm.

\subsection{Experimental results}

\begin{figure}
\centering \includegraphics[width=0.8\linewidth]{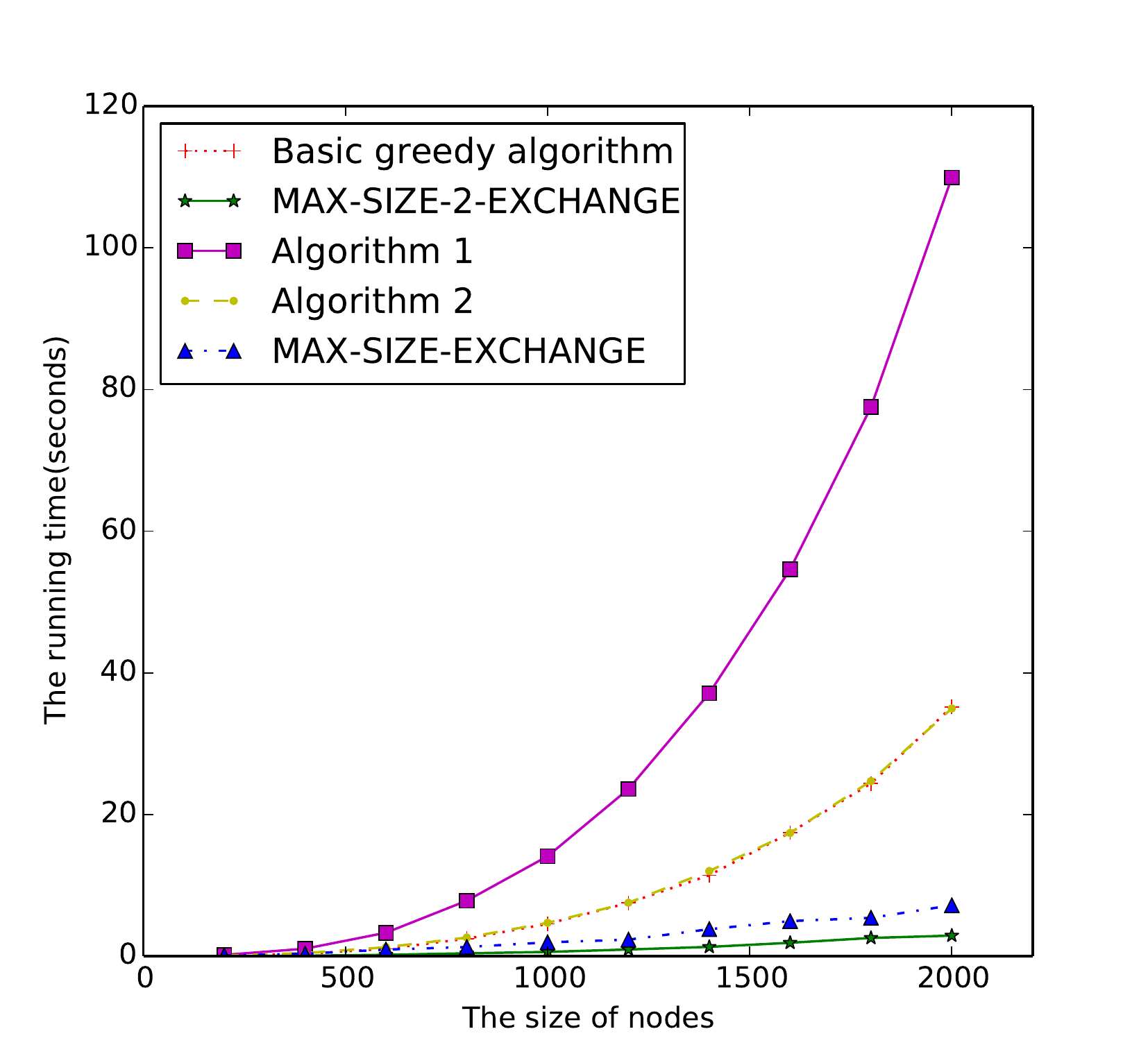}

\protect\caption{\label{fig:The running time of U.S. data}The running time of the
US data}
\end{figure}

\begin{figure}[t]
\centering \includegraphics[width=0.8\linewidth]{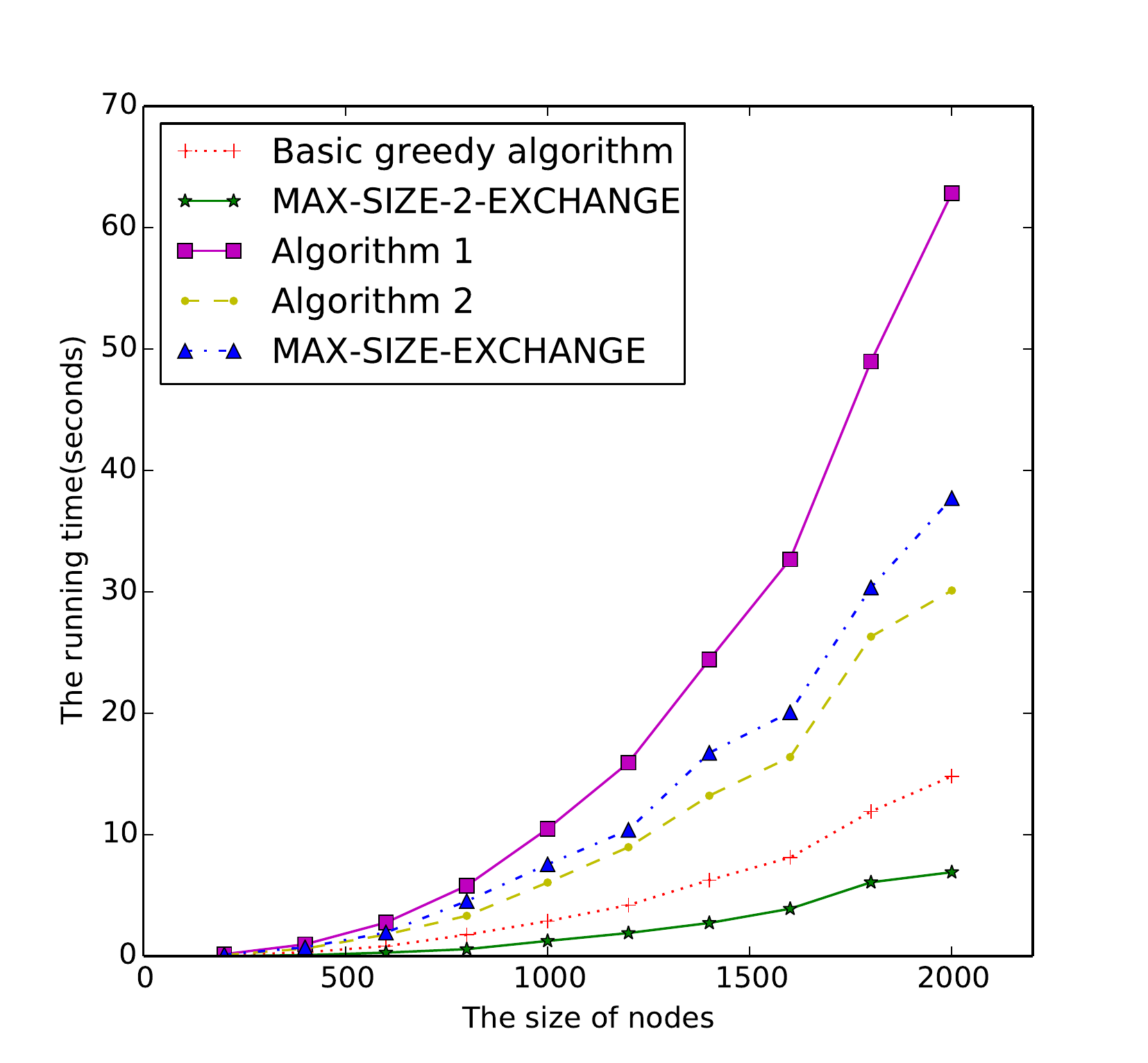}

\protect\caption{\label{fig:The running time of Chinese data}The running time of the
China data}
\end{figure}

Figure~\ref{fig:The experiment results of U.S. data} and \ref{fig:The experiment results of Chinese data}
show the quality of solutions for each size of graph in proportion
to \textsc{Max-Size-Exchange} in the US data and the China data respectively.
Figure~\ref{fig:The running time of U.S. data} and \ref{fig:The running time of Chinese data}
show the running time for each size of nodes in the US data and the
China data respectively.

The difference between the US data and China data is that the simulated
graph of the China data is sparser than the US data. This leads to
the result that the running time in the US data is more than in the
China data while the result of the US data is higher than the China
data. The approximation ratio of algorithm 1 is $3$, while algorithm
2 runs faster and yields better solution.

As shown in Figure~\ref{fig:The experiment results of U.S. data}
and~\ref{fig:The experiment results of Chinese data}, the solutions
of our algorithms are extremely close to the upper bound. In particular,
algorithm 2 has even figured out the optimal solution for $5$ times
out of $10$ test points in the case that the size of nodes is $2000$
in the US data. Both algorithms return within two minutes.

To sum up, both algorithm 1 and 2 are easy to implement and yield
good experimental performance on simulated data.

\section{Conclusion and future work}

We explore computational complexity of the clearing problem in the
kidney exchange market. Our inapproximability results, in comparison
to the best existing ones, are summarized in Table~\ref{tab:Summary-of-inapproximability}.
We have proposed two algorithms which run in polynomial time and perform
well on the simulated data as illustrated in Figure~\ref{fig:The experiment results of U.S. data}.
Both of these two algorithms are easy to implement, and give the solution
which are very close to the optimal. We make a conclusion that, the
kidney exchange problem in practice can be solved by a satisfactory
solution using the practical algorithms.

\begin{table}[b]
\centering \protect\caption{\label{tab:Summary-of-inapproximability}Summary of inapproximability
results.}

\begin{tabular}{|c|c|c|}
\hline 
Problem  & Our ratio  & Previous \tabularnewline
\hline 
\hline 
\textsc{Max-Weight-$L$-Exchange}  & $14/13$  & \textsc{APX-hard}\tabularnewline
\hline 
\textsc{Max-Weight-3-Exchange}  & $434/433$  & \textsc{APX-hard}\tabularnewline
\hline 
\textsc{Max-Size-3-Exchange}  & $698/697$  & \textsc{APX-hard}\tabularnewline
\hline 
\end{tabular}
\end{table}

There are several exciting directions for future research. First of
all, we are interested in closing the gap between approximation and
inapproximability. In particular, we are interested in improving the
$\tfrac{14}{13}$ result. In addition, we are interested in the complexity
of the clearing problems in similar matching markets. From a practical
perspective, to test the average performance, we are going to experimentally
test the new algorithms using real data.

\balance  \bibliographystyle{named}
\bibliography{ijcai15}

\end{document}